\documentstyle[aps,epsfig,prl]{revtex}
\begin{document}
\title{ Comment on ``Kagom{\'e} Lattice Antiferromagnet Stripped to Its Basics"}
\author{Steven R. White$^1$ and Rajiv R. P. Singh$^2$}
\address{$^1$Department of Physics, University of California, Irvine, CA 92697}
\address{$^2$Department of Physics, University of California, Davis, CA 95616}
\twocolumn[\hsize\textwidth\columnwidth\hsize\csname
@twocolumnfalse\endcsname
\maketitle
\widetext
\pacs{}

]

\narrowtext

In a recent letter, Azaria et al \cite{azaria} studied a
3-spin wide strip of the Kagome-lattice spin-$\frac{1}{2}$ Heisenberg model,
with the goal of understanding the large number of low-lying singlet
states observed in 2D Kagome clusters\cite{waldtmann}.
Using a number of approximate
field-theoretical mappings,  they concluded that this
system had a nondegenerate, undimerized ground state,
with a gap to spin excitations, but with gapless singlet
excitations. 
The Lieb-Schultz-Mattis theorem \cite{lsm}, which requires 
there to be at least one 
additional zero-energy state in the 
thermodynamic limit, allows this never-before-seen possibility.

A subsequent study \cite{pati}, using the numerical
density matrix renormalization group (DMRG)\cite{dmrg}, 
verified the existence
of a spin-gap, but was inconclusive about the key issues of degenerate
ground states and gapless singlet excitations.
Here, also using DMRG, we study much larger systems to examine 
these issues. We find that, contrary to the results of Azaria et al,
the ground state of this system is spontaneously dimerized, with
degenerate ground states. There is a very small spin-gap 
in the system but also a gap to singlet excitations. 
Above the ground states,
the gap to the singlet excitations is larger than
for the triplets. These results imply 
that this system is more analogous to the Majumdar-Ghosh 
model\cite{ghosh},
rather than to a novel spin liquid. 
Thus, the underlying field theory needs to be reexamined.

We studied systems up to length $1024\times3$, keeping up to 
400 states per block, using open boundary conditions. We found that 
the unmodified open ends of the strip have low lying triplet end
excitations, making it difficult to observe the bulk gaps. Therefore,
we terminated the ends using a $2\times2$ cluster of spins, as shown
on the left side of Fig. 1, which served to push all end excitations
above the bulk gaps.  Here,
all exchange couplings on the ends and in the bulk have identical
values $J$. 
In Fig. 2 we show the gap to the lowest lying triplet
state, with the modified ends, as a function of the system length. 
We are able to resolve
a very small triplet gap of $\Delta/J = 0.0104(5)$. Details of
the fit will be given elsewhere.

\begin{figure}[ht]
\epsfxsize=2.8 in\centerline{\epsffile{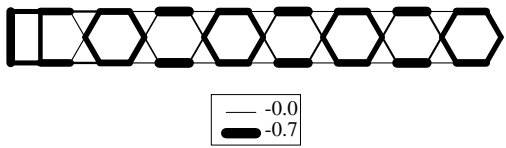}}
\caption{The local bond strengths 
$\langle \vec S_i \cdot \vec S_j \rangle$ 
for the left end of a $32\times 3$ strip are shown using 
the widths of the lines.
}
\end{figure}

\begin{figure}[ht]
\epsfxsize=2.0 in\centerline{\epsffile{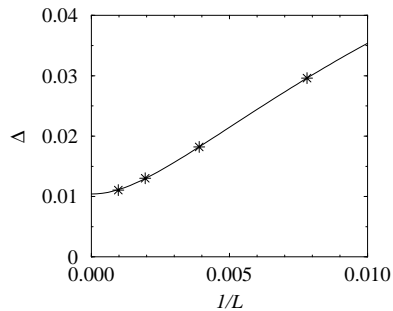}}
\caption{Singlet-triplet gap $\Delta$ for $L\times3$ strips as a function 
of the inverse length of the system. 
}
\end{figure}

We find that the bulk is dimerized. In Fig. 1, we show the local
bond strengths, with a clearly visible dimerization pattern,
on one end of a small $32\times3$ system.   Results for
systems as large as $1024\times3$ demonstrate that this dimerization
pattern persists in the bulk. For example, in the bulk we find
that the value of 
$\langle \vec S_i \cdot \vec S_j \rangle$ 
with $i$ and $j$ taking sequential values along the first leg 
follows the pattern: 
-0.071, -0.529, -0.071, -0.635, -0.071, -0.529, etc.
These values are well-converged both in the length of the system
and in the number of states kept. The singlet state representing
the shifted dimerization pattern ground state is visible
using periodic boundary conditions, where we found
a single very low lying singlet excited state below the triplet gap
on systems as large as $48\times3$.
In open systems, the boundaries push this state
above the triplet gap.  The entire pattern
of states is very similar to that of the Majumdar-Ghosh model.
These Kagome strips do not provide insight into the large number
of singlet
states observed in 2D Kagome clusters.

We thank Ian Affleck for discussions.
This work is supported in part by the NSF
under grants DMR98-70930, PHY94-07194, and DMR96-16574.

\end{document}